# Electron scattering near an itinerant to localized electronic transition

F. Rivadulla, J.-S. Zhou, and J. B. Goodenough

*Texas Materials Institute, ETC 9.102, The University of Texas at Austin, 1 University Station, C2201, Austin, TX 78712*

**Abstract**.- We report an unconventional temperature dependence of the resistivity in several strongly correlated systems approaching a localized to itinerant electronic transition from the itinerant electron side. The observed $(\rho-\rho_0) \propto T^{3/2}$ behavior over the entire range of materials discussed cannot be explained within the framework of existing theories. We propose a model in which the scattering of the conduction electrons by locally cooperative bond-length fluctuations in a matrix of vibronic and Fermi-liquid electrons can account for the experimental data.

INTRODUCTION

The transition from localized to itinerant electronic behavior in transition-metal compounds is characterized by a number of unusual physical properties, *e.g.* high-temperature superconductivity in layered copper oxides and a colossal magnetoresistance in manganese-oxide perovskites. Description of the evolution of the electrical resistivity on crossing this transition remains an open problem. On the localized-electron side of the transition, strong on-site electron-electron interactions separate successive redox energies by a finite energy gap, which makes single-valent compounds insulators; the electronic charge carriers of a mixed-valent system are polaronic, and polarons move diffusively with a mobility $\mu = (A'/T)\exp(-\Delta H_m/kT)$. The time $\tau_h$ for a polaron to transfer to a neighboring like atom on a crystallographically equivalent site is long compared to the period $\omega_o^{-1}$ of the locally cooperative optical-mode lattice vibrations that trap it in a potential well ($\tau_h > \omega_o^{-1}$). On the itinerant-electron side of the transition, a $\tau_h < \omega_o^{-1}$ allows the electrons to tunnel from site to site with a momentum $\hbar \boldsymbol{k}$ for a mean free time $\tau_s$ before it is scattered by a perturbation of the periodic potential; the electron mobility is $\mu = e\tau_s/m^*$, and the resistivity can be described by

$$\rho = \rho_o + AT^\alpha \qquad (1)$$

where the scaling parameter $\alpha$ depends on the scattering mechanism. In the conventional Fermi-Liquid (FL) framework, a coulombic electron-electron repulsion at very low temperatures gives a value of $\alpha = 2$ while electron-phonon effects dominate at higher temperatures ($T > \Theta_D/2$) to give an $\alpha = 1$ [1].

Brinkman and Rice [2] have shown that on the approach from the itinerant electron side to crossover to localized-electron behavior, the effective mass of an itinerant electron is enhanced as

$$m_{ee}^* = m_b/[1-(U/U_c)^2] \qquad (2)$$

where U is the on-site coulomb energy to transfer an electron from one atom to a like near neighbor and $m_b$ is the bare-electron mass. However, on closer approach to the Mott-Hubbard limit $U=U_c$, experiment has shown directly [3] and indirectly [4, 5] the appearance of strong-correlation fluctuations in an itinerant-electron matrix in several single-valent transition-metal oxoperovskites. The infinite effective mass projected by eq. (2) is avoided by introducing phase fluctuations that are created by locally cooperative (M-O) bond-length fluctuations in the transition-metal $MO_3$ array of an $AMO_3$ perovskite. Strong electron coupling to bond-length fluctuations in a mixed-valent $MnO_3$ array has been shown [6] to result in a vibronic diffusive conduction having a mobility $\mu = eD_o/kT$ in which $D_o \propto \omega_o$ rather than the attempt frequency $v_o > \omega_o$; vibronic conduction gives "bad-metal" behavior having an $\alpha = 1$ dependence in eq. (1), but without the saturation of $\rho(T)$ at high temperatures that is characteristic of the Boltzmann equation for FL behavior [1].

The breakdown of conventional electron-phonon scattering within FL theory has been extensively discussed for the case of optimally doped and overdoped superconductive cuprates, but a general consensus as to the appropriate physical model of the phenomenon has not been reached [7]. A non-FL temperature dependence of $\rho(T)$ has recently been reported for several other metals previously assumed to be conventional itinerant metals. Notable examples are the paramagnetic $\beta$ phase of elemental Mn [8] and the high-pressure phase of the itinerant ferromagnet MnSi [9]. Self-consistent renormalization (SCR) theory [10], and other variations of the FL model have successfully predicted the $\rho(T)$ behavior of several narrow-band itinerant-electron systems like $NiS_{2-x}Se_x$ [11], $ZrZn_2$ [12], and some heavy fermion compounds, but they have failed to describe the $\rho(T)$ data reported for MnSi, $La_4Ru_6O_{19}$ [13] or the critical properties of the $CeCu_{6-x}Au_x$ system [12].

We report an unconventional $\alpha = 3/2$ in eq. (1) for the hole-doped $Na_{0.5}CoO_2$ and the pressure-induced metallic phase of electron-doped $BaCoS_2$. Moreover, we show that the observed behavior is common to other strongly correlated systems that approach the localized-itinerant electronic transition from the itinerant-electron side. We argue that the $\alpha = 3/2$ is a general phenomenon found in metals that approach the Mott-Hubbard transition from the itinerant-electron side and that existing *ad hoc* models do not embrace this generality. We present a model of electron scattering from locally cooperative bond-length fluctuations that should be generally applicable to a narrow bandwidth interval between long-range magnetic order and FL behavior for both single-valent and mixed-valent transition-metal compounds.

**EXPERIMENTAL**

Single crystals of $BaCo_{0.9}Ni_{0.1}S_{2-x}Cl_x$ ($0 \leq x \leq 0.17$) were synthesized in evacuated silica tubes from polycrystalline samples [14]. The polycrystalline materials were prepared by reaction of BaS, Co, Ni, S and $BaCl_2$ inside evacuated silica tubes at 920ºC for three days with an intermediate grinding. The material was quenched to an ice-water bath from 920ºC.

$BaCoS_2$ has a layered structure of CoS planes separated by rock-salt BaS layers [15]. An antiferromagnetic Néel temperature $T_N$ = 305 K is progressively lowered by doping the CoS planes with electrons in $BaCo_{1-x}Ni_xS_2$, a semiconductor to metal transition occurring at x=0.22 where $T_N$ falls to zero [16]. An anomalous electronic behavior near the transition [17] stimulated a search for superconductivity in this system, but none has been induced by doping and/or pressure. It was pointed out that a strong Co:3d-S:3p hybridization with the apical sulfur atoms gives the system a quasi-3D electronic character, which was suggested to be a principal cause for the absence of superconductivity [18]. On the other hand, neutron diffraction has shown that the insulating phase contains high-spin S = 3/2 Co(II), and the high-spin cobalt of a sulfur deficient $BaCo_{0.9}Ni_{0.1}S_{1.87}$ transforms at a critical pressure $P_c$ to a low-spin-cobalt, high-pressure metallic phase [14]. We chose to achieve the additional electron doping introduced by sulfur vacancies with a substitution of Cl for S. The evolution with x of lattice parameters and $T_N$ correspond to those for $BaCo_{1-x}Ni_xS_2$, which confirms that Cl has substituted for S to introduce electrons in the CoS planes. We assume that $Cl^-$ ions substitute in the BaS layer where they are apical with respect to the Co atom coordination, but this point should be confirmed by neutron diffraction. Since we could not induce a metallic phase with Cl doping, we used high-pressure to reach the crossover. Resistivity under pressure was measured by the four-probe method in a Be-Cu cell, up to 20 kbar.

Removal of Na from the layered $NaCoO_{2\pm\delta}$ produces a change from polaronic to metallic resistivity [19]. $Na_{0.5}CoO_2$ is on the metallic side of the transition, which makes it a good candidate for studying transport properties at the approach to the transition to localized electronic behavior. Single crystals of $Na_{0.5}CoO_2$ were synthesized from a NaCl flux containing $Co_3O_4$ and $Na_2CO_3$ in a molar ratio Co:Na:NaCl=1:1:7. After heating the mixture at 950ºC for 12 h, the temperature was slowly cooled down to 800ºC (0.5ºC/h) and then at 180ºC/h down to room temperature. Single crystals of typical dimensions $1 \times 1 \times 0{,}02$ mm$^3$ were obtained.

**RESULTS**

Fig. 1 shows the temperature dependence of the resistivity in the *ab* plane of a BaCo$_{0.9}$Ni$_{0.1}$S$_{1.85}$Cl$_{0.15}$ single crystal under different pressures. At atmospheric pressure, the high-spin phase is an antiferromagnetic semiconductor with a $T_N \approx 225$ K. The $\rho(T)$ curve shows a change of slope at $T_N$. In the interval $6.4 < P < 7.8$ kbar, the high-temperature slope changes to metallic and a thermal hysteresis is associated with a polaronic-metallic transition at $T_{PM}$ for pressures $5.6 \leq P \leq 8.9$ kbar. This pressure dependence is similar to that found [14] for single-crystal BaCo$_{0.9}$Ni$_{0.1}$S$_{1.87}$. The smooth transition to metallic conductivity at $T_{PM}$ despite a first-order character signals the coexistence of the antiferromagnetic, polaronic phase and the metallic phase with a growth under pressure of the volume fraction of the metallic phase to beyond percolation for $P > 9.6$ kbar.

Fig. 2 shows the variation of $T_{PM}$ as determined from the change of slope of $\rho(T)$ with a thermal hysteresis at $T_{PM}$, but not at $T_N$. A pressure hysteresis as well as a thermal hysteresis is associated with the transition at $T_{PM}$.

Fig. 3 shows that the metallic phase at 12.6 kbar of BaCo$_{0.9}$Ni$_{0.1}$S$_{1.85}$Cl$_{0.15}$ has a $\rho_{ab}(T)$ described by eq. (1) with a scaling parameter $\alpha = 1.5$ over the entire temperature range $3$ K $< T < 300$ K. A conventional metal would have an $\alpha = 1$ at intermediate temperatures, and a strongly correlated FL would have a $\alpha = 2$ at low temperatures. At very low temperatures where coulombic repulsion should dominate the resistivity, experimental problems can introduce non-intrinsic artefacts into the measurement. However, conventional electron-phonon scattering, which has linear temperature dependence, is clearly replaced at higher temperatures by some other scattering mechanism.

Fig. 4 shows the variation of $\alpha$ for $\rho_{ab}(T)$ of BaCo$_{0.9}$Ni$_{0.1}$S$_{1.85}$Cl$_{0.15}$ in the metallic temperature range $T > T_{PM}$. The dashed lines mark the pressure interval $\Delta P$ over which $T_{PM}$ is found with increasing pressure; the thermal hysteresis associated with $T_{PM}$ disappears with decreasing pressure below 7 kbar. In the inteval $7 \leq P \leq 9.2$ kbar with decreasing pressure, the metallic phase at T > $T_{PM}$ has a $\rho_{ab}(T)$ described by eq. (1) with an $\alpha$ that decreases with decreasing pressure from $\alpha \approx 3/2$ at 9.2 kbar to $\alpha \approx 1.35$ at 7 kbar.

Fig. 5 shows that $\rho_{ab}(T)$ for $Na_{0.5}CoO_2$ also follows a $T^{3/2}$ law over a wide temperature range. The deviation from a $T^{3/2}$ dependence at lowest temperatures may not be intrinsic [20]. At $\approx$ 180 K, the material becomes semiconductor along the *c*-axis [21] and deviates progressively from the $T^{3/2}$ law above this temperature.

The perovskite $PrNiO_3$ exhibits a first-order transition from AF insulator to metallic behavior at a $T_N = T_t = 125$ K that decreases with increasing pressure [22]. Moreover, the pressure experiments have shown a progressive change to bad-metal behavior below $T_N = T_t$ at a pressure $P_c$ where $T_N = T_t$ is still finite, which is indicative of the coexistence of the metallic and AF phases with the metallic phase passing a percolation threshold at $P = P_c$ [22]. The lower inset of Fig. 5 shows that the 3D conduction at $P > P_c$ follows a $T^{3/2}$ law over a huge temperature range.

Direct observation of the coexistence of strongly correlated and itinerant electrons has been made with photoemission spectroscopy in the perovskite systems $Ca_{1-x}Sr_xVO_3$, $La_{1-x}Sr_xTiO_3$, and $La_{1-x}Sr_xVO_3$ [3,7]. The first of these approaches $U = U_c$ as x decreases to x = 0 and the two mixed-valent systems undergo a transition from antiferromagnetic insulator to bad-meal behavior with increasing x. As can be seen in the upper inset of Fig. 5, $CaVO_3$ also obeys the $T^{3/2}$ law from room temperature down to $\approx$ 80 K. Below this temperature, $\rho(T)$ fits a $T^2$ law. Allen [1] has called attention to a similar "non-conventional" metallic behavior for the other end member of this system, $SrVO_3$.

Of particular interest is $La_{1-x}Sr_xVO_3$, which is metallic with a $T^{3/2}$ dependence for x > 0.22 where $T_N$ is totally suppressed [23].

**DISCUSSION**

It is apparent from these examples that a number of compounds exist where the resistivity can be described by eq. (1), but with a scaling parameter $\alpha = 3/2$, or between 1 and 3/2; this temperature dependence is at variance with conventional itinerant-electron theories for a single-phase system. Moreover, most of these share as a common property the approach to localized-electron behavior from the itinerant-electron side, an approach where strong-correlation fluctuations in an itinerant-electron matrix have been found [6]. The first question to ask is whether existing *ad hoc* models can be extended to embrace the entire range of compositions where this phenomenon is found. If not, a more general model needs to be considered.

Frozen disorder in amorphous metals and spin glasses can produce a resistivity described by eq. (1) with $\alpha = 3/2$ [24]. However, this mechanism is only operative at very low temperatures whereas a $T^{3/2}$ temperature dependence to 300 K was found in the examples reported in Fig. 3 and 5. Moreover, in some of the materials discussed below, atomic disorder is not an issue, which rules out this interpretation.

On the other hand, SCR theory [10] extends the FL model to include coupling of the conduction electrons to spin fluctuations. This theory predicts different values of the parameter $\alpha$ in eq. (1) depending on the sign of the spin-spin interaction and the dimensionality of the system. For example, $\alpha = 3/2$ is predicted for a 3D itinerant system coupled to antiferromagnetic (AF) spin fluctuations. For this model to be applicable to the systems we are dealing with here, strong AF spin fluctuations must produce a Curie-Weiss temperature dependence of $\chi$. Unfortunately we are unable to measure $\chi(T)$ under high pressure, but a comparison with $\chi$ of the $BaCo_{1-x}Ni_xS_2$ system indicates almost no deviation of $\chi(T)$ from a temperature-independent Pauli paramagnetism in metallic samples with $x > 0.25$ [25].

LSDA calculations [26] showed that $Na_{0.5}CoO_2$ presents predominantly ferromagnetic spin fluctuations, which again makes it impossible to interpret the $(\rho-\rho_0) \propto T^{3/2}$ dependence in the framework of SCR theory.

In $La_{1-x}Sr_xVO_3$, the failure of the NMR experiments [27] to find any evidence of AF spin fluctuations in the metallic phase despite a $(\rho-\rho_0) \propto T^{3/2}$ dependence shows that the SCR theory is not applicable to this systems either. In addition, it has already been pointed out [28] that SCR theory is not applicable to the $LnNiO_3$ family as it predicts a Curie constant that is independent of $T_N$, which is contrary to experiment. On the other hand, strong correlation fluctuations carry a spin, and the spin fluctuations may add a $T^{3/2}$ contribution to the resistivity.

The failure of existing models to account for the $\rho(T)$ data on the approach to the transition to localized-electron behavior from the itinerant-electron side has prompted us to consider first the evolution of the electronic states on crossing the Mott-Hubbard transition from the itinerant-electron side. All the existing models to describe the transition assume a homogeneous electronic system [7]. As we illustrated schematically in Fig. 6, the crossover at a first-order Mott-Hubbard transition begins with the introduction of strong-correlation fluctuations in a FL matrix. The equilibrium cation-anion bond length is longer within a strong-correlation fluctuation than that of the matrix. Therefore, the coexistence of strongly correlated electrons in a lower Hubbard band and FL electrons having a well-defined Fermi surface perturbs the periodic potential so as to introduce Anderson-localized states below a lower mobility edge $\mu_c$ and above an upper mobility edge $\mu_c'$. But in this case, the perturbation of the periodic potential is not static, so the Anderson-localized states become vibronic states. The FL states are confined to the interval $\mu_c < E < \mu_c'$, and as the volume of the strong-correlation fluctuations increases, the energy interval $\mu_c' - \mu_c$ decreases until all the states of the vibronic/FL band become vibronic as is proposed in the normal state of the copper-oxide superconductors [29]. Complete crossover to an insulator in a single-valent system or to a polaronic conductor in a mixed-valent system would proceed by isolating mobile vibronic clusters within a localized-electron matrix with the volume fraction of vibronic clusters disappearing in a single-valent compound or collapsing to small polarons in a mixed-valent system. Our focus is on systems having a coexistence of strong-correlation fluctuations in a matrix of vibronic/FL states. In particular, we are interested in cases where the Fermi energy $E_F$ lies close to a mobility edge $\mu_c$ or $\mu_c'$, a condition that can be satisfied in a mixed-valent system or a single-valent compound with an orbital degeneracy. For this situation, we can assume that the contribution to the conductivity of the FL electrons dominates that of the vibronic electrons and that the Fermi-Dirac distribution function $f(E)$ for the FL electrons can be replaced by a Boltzmann distribution function so that $\partial f(E)/\partial E \approx (kT)^{-1} f(E)$. With these two assumptions, calculation of the mean scattering time of the FL electrons is similar to that for scattering of electronic charge carriers by lattice vibrations in semiconductors [30].

The scattering probability in this situation is proportional to

$$\left| \sum_{ij} \Delta_{ij} E_{ij} \right|^2 N(E) \qquad (3)$$

where $\Delta_{ij}$ is the amplitude of a component of the local, thermally fluctuating dilatation and $E_{ij}$ is a component of the deformation-potential tensor. The mean square density fluctuation is

$$|\overline{\Delta}|^2 \propto kT \qquad (4)$$

Because the density of states for a 3D isotropic gas of itinerant electrons is

$$N(E) = \frac{1}{2\pi^2}\left(\frac{2m^*}{\hbar^2}\right)^{3/2} E^{1/2} \qquad (5)$$

the relaxation time, which is proportional to the inverse of (3), becomes

$$\tau(E) \propto T^{-1} E^{-1/2} m^{*-3/2} \qquad (6)$$

The mean relaxation time can be calculated by substituting (5) and (6) into the general expression

$$\overline{\tau} = \frac{2}{3kT} \frac{\int E\tau(E)f(E)N(E)dE}{\int f(E)N(E)dE} \qquad (7)$$

which results in

$$(\rho-\rho_0) \propto (1/\overline{\tau}) \propto T^{3/2} \qquad (8)$$

for conduction in 3D systems. Although the conductivities of $BaCo_{0.9}Ni_{0.1}S_{1.85}Cl_{0.15}$ and $Na_{0.5}CoO_2$ are strongly anisotropic, nevertheless they are 3D (in the sense that the temperature dependence of the resistivity is nearly the same in all directions) in the temperature range where the resistivity is described by eq. (8). The validity of this approximation was demonstrated by Valla *et al.*[21] through a comparison of ARPES and conductivity results in the anisotropic systems $Na_{0.5}CoO_2$ and $(Bi_{0.5}Pb_{0.5})_2Ba_3Co_2O_y$. Although the anisotropy in the conductivity implies different values of m* in the *ab* plane and the *c* direction, the same energy dependence ($\propto E^{1/2}$) will hold in eq.(5), so long as the conductivity is metallic in 3D [30]. The fact that ρ(T) in Fig. 5 deviates progressively from the $T^{3/2}$ law above 180 K where $Na_{0.5}CoO_2$ becomes semiconductor in the *c* direction supports this contention.

Therefore, we conclude that the existence of locally cooperative bond-length fluctuations at isolated clusters is able to account for the experimentally observed $(\rho-\rho_0) \propto T^{3/2}$ in compounds that approach the Mott-Hubbard localized to itinerant electronic transition from the itinerant-electron side. Although spin fluctuations are normally associated with strong-correlation fluctuations [6] the model is not restricted to AF spin fluctuations.

Finally, we note that vibronic electrons, which move diffusively with a motional enthalpy $\Delta H_m < kT$, give a $(\rho-\rho_0) \propto T$ as do FL electrons at temperatures $T > \Theta_D/2$. Therefore $(\rho-\rho_0) \propto T^{3/2}$ refers to an intermediate situation, and we can expect an evolution of $(\rho-\rho_0) \propto T^\alpha$ at higher temperatures from $\alpha = 1$ to $\alpha = 3/2$ and back to $\alpha = 1$ on crossing from FL to vibronic conduction in 3D. Such an evolution appears to occur in $BaCo_{0.9}Ni_{0.1}S_{1.85}Cl_{0.15}$ on lowering the pressure, which increases the vibronic component of the conduction. We have also observed an $\alpha = 1.3$ in $LaNiO_3$ where the strong-correlation fluctuations are reduced relative to the situation in metallic $PrNiO_3$.

**CONCLUSIONS**

We have reported a $(\rho-\rho_0) \propto T^{3/2}$ temperature dependence of the resistivity over a large temperature range for several metallic systems that approach the first-order transition to localized-electron behavior from the itinerant-electron side. We have cited direct and indirect evidence for the coexistence of strong-correlation fluctuations containing electrons in a lower Hubbard band coexisting with FL electrons. We have argued that the itinerant-electron band of FL states contains vibronic states at the top and bottom of the band as a result of the perturbation of the periodic potential by the strong-correlation fluctuations, which have a larger equilibrium cation-anion bond length than the itinerant-electron matrix. We have also argued that where the Fermi energy lies close to a mobility edge in the vibronic/FL band, a $(\rho-\rho_0) \propto T^{3/2}$ is described by scattering of the FL electrons form the bond-length fluctuations. We hope this discussion will stimulate a more detailed theoretical consideration of the evolution of the temperature dependence of the resistivity on crossing from itinerant to localized electronic behavior.

**Acknowledgments**

We would like to thank Dr. J. Fernandez-Rossier for fruitful discussions. We also acknowledge support from the NSF, the Robert A. Welch Foundation, Houston TX, and the Texas Center for Superconductivity at the University of Houston (TCSUH). F. R. would like to thank the Fulbright foundation and the Ministery of Science of Spain for a postdoctoral fellowship.

**REFERENCES**

1 P. B. Allen, Comments Cond. Mat. Phys. **15**, 327 (1992).

2 W. F. Brinkman, and T. M. Rice, Phys. Rev. **2**, 4302 (1970).

3 I. H. Inoue, I. Hase, Y. Aiura, A. Fujimori, Y. Haruyama, T. Maruyama, and Y. Nishihara, Phys. Rev. Lett. **74**, 2539 (1995).

4 J.-S. Zhou, and J. B. Goodenough, Phys. Rev. B **54**, 13393 (1995).

5 I. H. Inoue, O. Goto, H. Makino, N. E. Hussey, and M. Ishikawa, Phys. Rev. B **58**, 4372 (1998); H. Makino, I. H. Inoue, M. J. Rozemberg, I. Hase, Y. Aiura, S. Onari, Phys. Rev. B **58**, 4384 (1998).

6 J. B. Goodenough and J.-S. Zhou, in *Structure and Bonding*, vol 98, Ed. by J. B. Goodenough, Springer-Verlag, Berlin 2001.

7 M. Imada, A. Fujimori, Y. Tokura, Rev. Mod. Phys. **70**, 1039 (1998).

8 J. R. Stewart, B. D. Rainford, T. S. Eccleston, and R. Cywinski, Phys. Rev. Lett. **89**, 186403 (2002).

9 C. Pfleiderer, S. R. Julian, and G. G. Lonzarich, Nature **414**, 427 (2001).

10 T. Moriya, Spin Fluctuations in Itinerant Electron Systems, Springer (Berlin, 1985).

11 M. Kamada, N. Mori, and T. Mitsui, J. Phys. C.: Solid State Phys. **10**, L643 (1977).

12 C. M. Varma, Z. Nussinov, and W. van Saarloos, Phys. Rep. **361** 267 (2002).

13 P. Khalifah, K. D. Nelson, R. Jin, Z. Q. Mao, Y. Liu, Q. Huang, X. P. A. Gao, A. P. Ramirez, and R. J. Cava, Nature **411**, 669 (2001).

14 J.-S. Zhou, W. J. Zhu, and J. B. Goodenough, Phys. Rev. B **64**, 140101 (2001).

15 I. E. Grey, and H. Steinfink, J. Am. Chem. Soc. **92**, 5093 (1970).

16 L. S. Martinson, J. W. Schweitzer, and N. C. Baenziger, Phys. Rev. Lett. **71**, 125 (1993).

17 T. Sato, H. Kumigashira, D. Ionel, T. Takahashi, I. Hase, H. Ding, J. C. Campuzano, and S. Shamoto, Phys. Rev. B **64**, 075103 (2001).

18 K. Takenaka, S. Kashima, A. Osuka, S. Sugai, Y. Yasui, S. Shamamoto, and M. Sato, Phys. Rev. B **63**, 115113 (2001).

19 S. Kikkawa, S. Miyazaki, and ; Koizumi, J. Solid State Chem. **62**, 35 (1986).

20 Due to the very small dimensions of the crystal of $Na_{0.5}CoO_2$ ($1.5\times1.5\times0.04$ mm$^3$) and the high current used (3 mA) in order to get a good signal to noise ratio, the effective temperature of the sample could be higher than the one read by the thermocouple. At low temperatures, this effect may not be negligible and could be responsible for the deviations from the T3/2 law observed in the main panel of fig. 5 below ≈20 K.

21 T. Valla, P. D. Jonson, Z. Yusof, B. Wells, Q. Li, S. M. Loureiro, R. J. Cava, M. Mikami, Y. Mori, M. Yoshimura, and T. Sasaki, Nature **417**, 627 (2002).

22 J.-S. Zhou, J. B. Goodenough, B. Dabrowski, P. W. Klamut, and Z. Bukowski, Phys. Rev. B **61**, 4401 (2000).

23 F. Inaba, T. Arima, T. Ishikawa, T. Katsufuji, and Y. Tokura, Phys. Rev. B **52**, 2221 (1995).

24 N. Rivier, and K. Adkins, J. Phys. F: Metal Phys., **5,** 1745 (1975).

25 J. Takeda, Y. Kobayashi, K. Kodama, H. Harashina, and M. Sato, J. Phys. Soc. Japan **64**, 2550 (1995).

26 D. J. Singh, Phys. Rev. B **61**, 13397 (2000).

27 A. V. Mahajan, D. C. Johnston, D. R. Torgeson, and F. Borsa, Phys. Rev. B **46**, 10973 (1992).

28 J.-S. Zhou, J. B. Goodenough, B. Dabrowski, P. W. Klamut, and Z. Bukowski, Phys. Rev. Lett. **84**, 526 (2000).

29 J. B. Goodenough, Europhys. Lett. **57**, 550 (2002).

30 J. M. Ziman, *Electrons and Phonons*, Oxford, 1960.

**Figures & Captions**

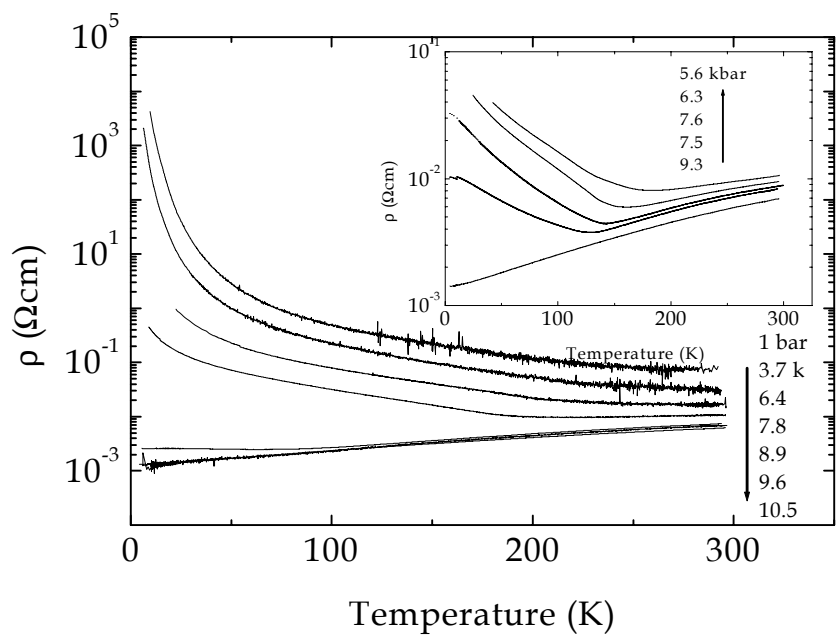

FIGURE 1

**Figure 1.-** Temperature dependence of the *ab* plane resistivity of a $BaCo_{0.9}Ni_{0.1}S_{1.85}Cl_{0.15}$ single crystal at different pressures. A semiconductor to metal transition is progressively reached as the pressure increases. Inset: recovery of the low temperature semiconductive state as the pressure is progressively released.

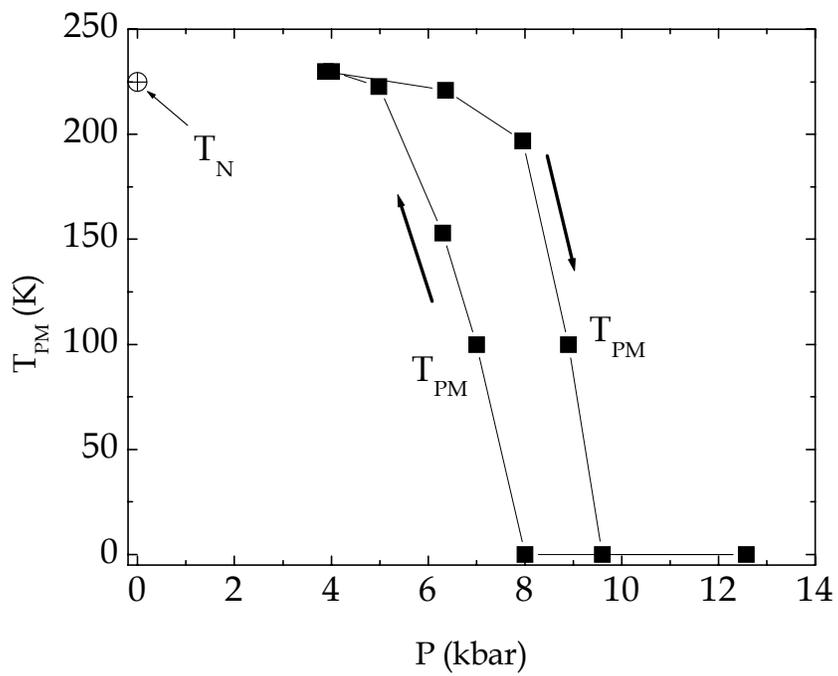

FIGURE 2

**Figure 2.-** Pressure dependence of the polaronic-metallic transition temperature, $T_{PM}$, of $BaCo_{0.9}Ni_{0.1}S_{1.85}Cl_{0.15}$.

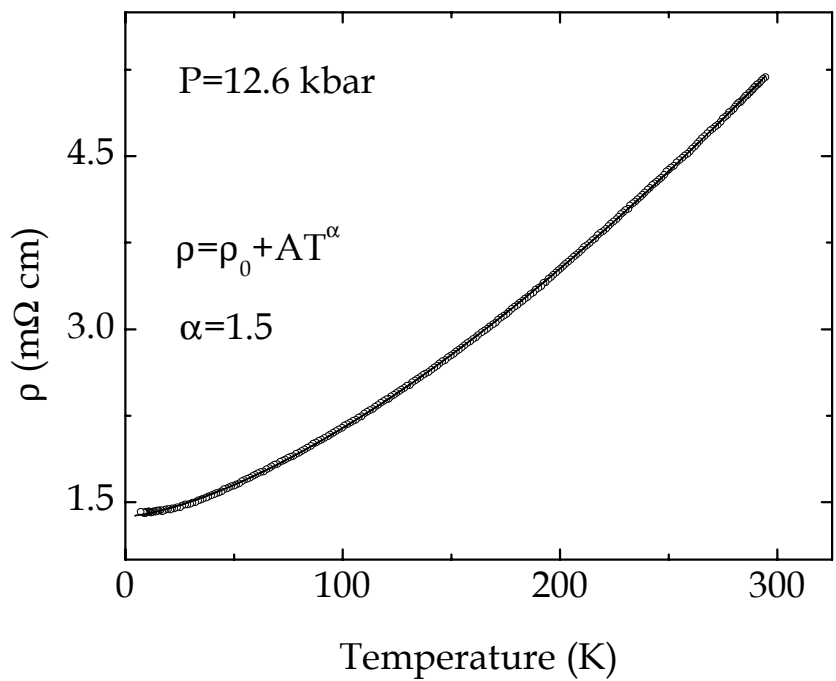

FIGURE 3

**Figure 3.-** Temperature dependence of the *ab* plane resistivity of a BaCo$_{0.9}$Ni$_{0.1}$S$_{1.85}$Cl$_{0.15}$ single crystal at 12.6 kbar (circles), along with the fit to eq. (1) with $\alpha = 3/2$ (line).

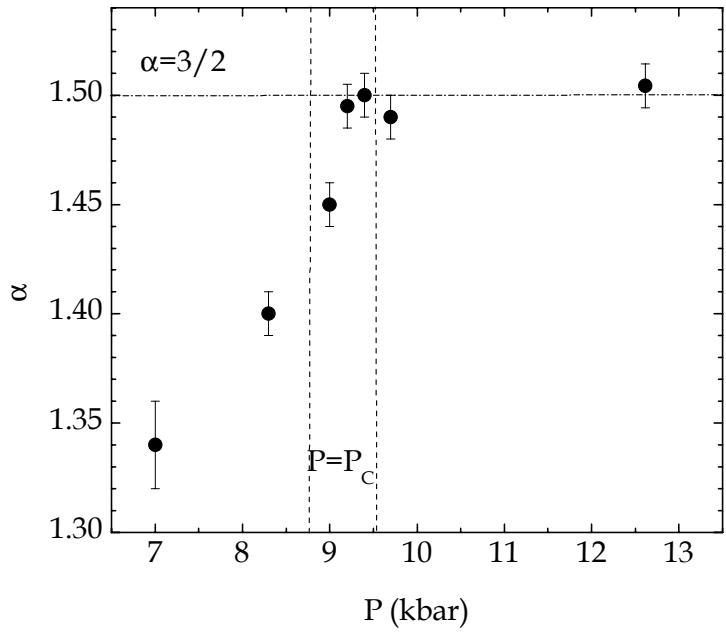

FIGURE 4

**Figure 4.-** Evolution of the exponent α from eq. (1) as a function of decreasing pressure for $\rho_{ab}$ of $BaCo_{0.9}Ni_{0.1}S_{1.85}Cl_{0.15}$. The vertical lines mark the pressure interval in which $T_{PM}$ is observed with increasing pressure.

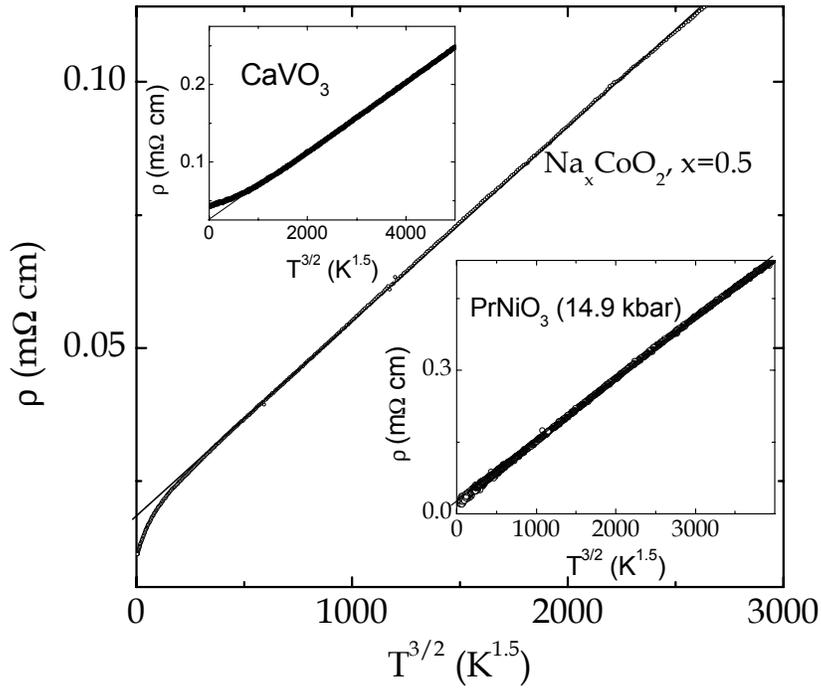

FIGURE 5

**Figure 5.-** The resistivity of different metallic materials approaching the localized transition, vs. $T^{3/2}$. Main panel: ab plane of $Na_{0.5}CoO_2$. Top inset: polycrystalline cold-press (high density) $CaVO_3$. In this case the resistivity becomes proportional to $T^2$ below $\approx 80$ K, which is also at odds with conventional models. Lower inset: polycrystalline $PrNiO_3$ (sintered at high oxygen pressure) measured at 14.9 kbar.

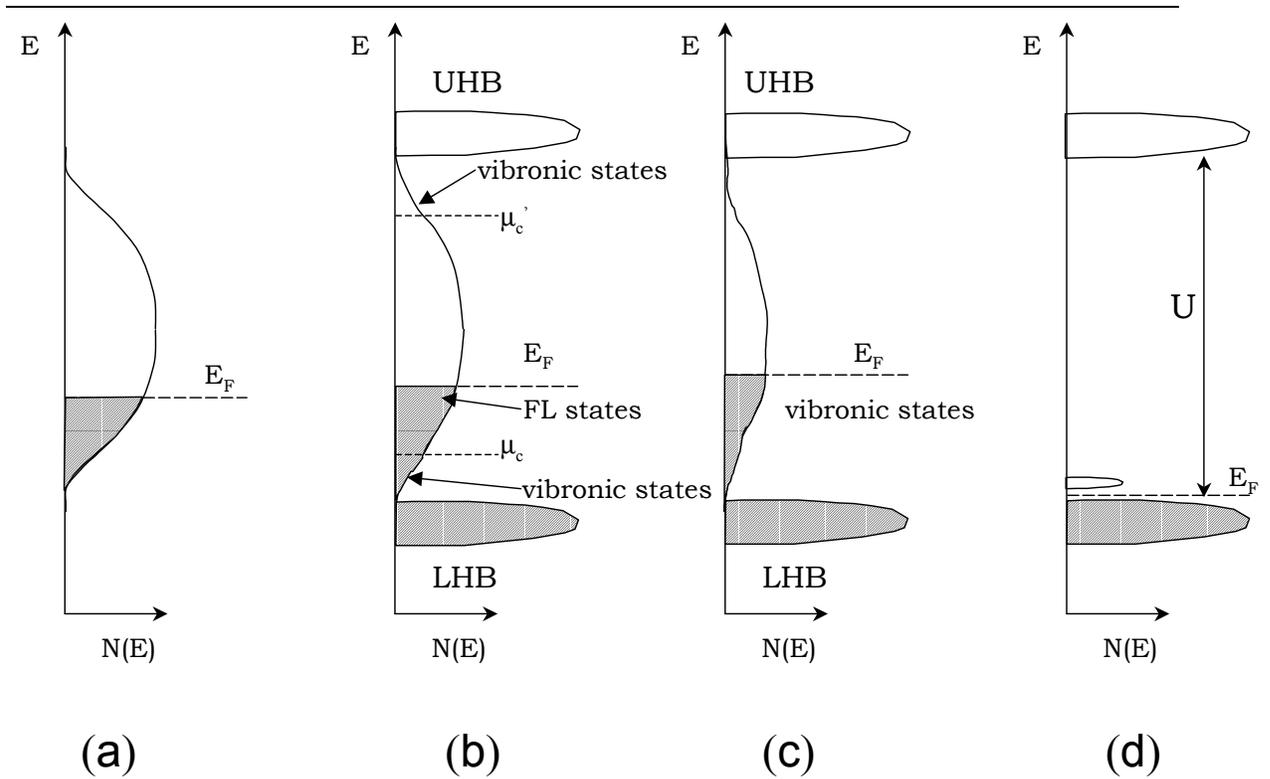

FIGURE 6

**Figure 6.-** Evolution of energy density of one-electron states for a non degenerate band at a crossover from FL to localized-electron behavior: (a) FL phase, (b) strong-correlation fluctuations in a FL matrix with vibronic states above or below a mobility edge at $\mu_c$, (c) vibronic phase, (d) polaronic phase for a p-type conduction. LHB and UHB refer to lower and upper Hubbard bands.